\documentclass{article}
\usepackage{spconf,amsmath,graphicx}
\usepackage{subcaption}
\usepackage{url}
\renewcommand{\paragraph}[1]{\noindent\textbf{#1}\quad}

\usepackage{enumitem}
\setlist{nosep, leftmargin=14pt}

\usepackage{mwe}

\title{Bottom-Up instance segmentation of catheters for Chest X-rays}

\name{Francesca Boccardi$^{\star \dagger}$ , Axel Saalbach$^{\star}$, Heinrich Schulz$^{\star}$, Samuele Salti$^{\dagger}$, Ilyas Sirazitdinov$^{\star}$ 
\thanks{This work has been submitted to the IEEE for possible publication. Copyright may be transferred without notice, after which this version may no longer be accessible.}
}
\address{
$^{\star}$ Philips Research Hamburg, Germany\\
$^{\dagger}$ University
of Bologna, Bologna, Italy\\
}
\begin{document}
\maketitle

\begin{abstract}
Chest X-ray (CXR) is frequently employed in emergency departments and intensive care units to verify the proper placement of central lines and tubes and to rule out related complications. 
The automation of the X-ray reading process can be a valuable support tool for non-specialist technicians and minimize reporting delays due to non-availability of experts.
While existing solutions for automated catheter segmentation and malposition detection show promising results, the disentanglement of individual catheters remains an open challenge, especially in complex cases where multiple devices appear superimposed in the X-ray projection.
Moreover, conventional top-down instance segmentation methods are ineffective on such thin and long devices, that often extend through the entire image.
In this paper, we propose a deep learning approach based on associative embeddings for catheter instance segmentation, able to overcome those limitations and effectively handle device intersections.

\end{abstract}


\section{Introduction}
\label{sec:intro}
Chest X-ray (CXR) is widely used to detect lung abnormalities and pathologies. In emergency departments and intensive care units, they are often employed to check the correct placement of medical devices in order to exclude potentially dangerous complications \cite{gambato2023chest}. In those cases, a careful assessment of the CXR is required, which is typically performed by the attending radiologist. The development of an automated process could assist non-professional technicians in the interpretation of X-rays when an expert is not available, potentially reducing reporting delays. Several efforts have been made to deploy automated solutions for catheter segmentation \cite{subramanian2019automated}, malposition recognition \cite{hansen2021radiographic}, and catheter tip detection \cite{jung2022classification}.
In addition, it is not rare that a single CXR may contain multiple devices, often appearing as superimposed in the projection image. In such scenarios, it is essential to identify and separate individual catheters. Conventional methods for instance segmentation, like Mask R-CNN \cite{he2017mask}, rely on a detect-and-segment strategy and may not be effective for this specific application. Indeed, the typically low resolution of the instance segmentation branches does not enable an accurate segmentation of thin and long structures such as catheters. To overcome these limitations, we propose a deep-learning approach based on associative embedding for cardiac catheter instance segmentation. The key contributions of this study can be summarized as follows: adaptation of the LaneNet \cite{neven2018towards} architecture for catheter instance segmentation,
by using HRNet(V2) \cite{sun2019high} as a customized backbone; an intersection resolution algorithm to effectively handle catheters intersections.

\section{Methodology}
\label{sec:methods}

\begin{figure*}[t!]
\includegraphics[width=\linewidth, keepaspectratio]{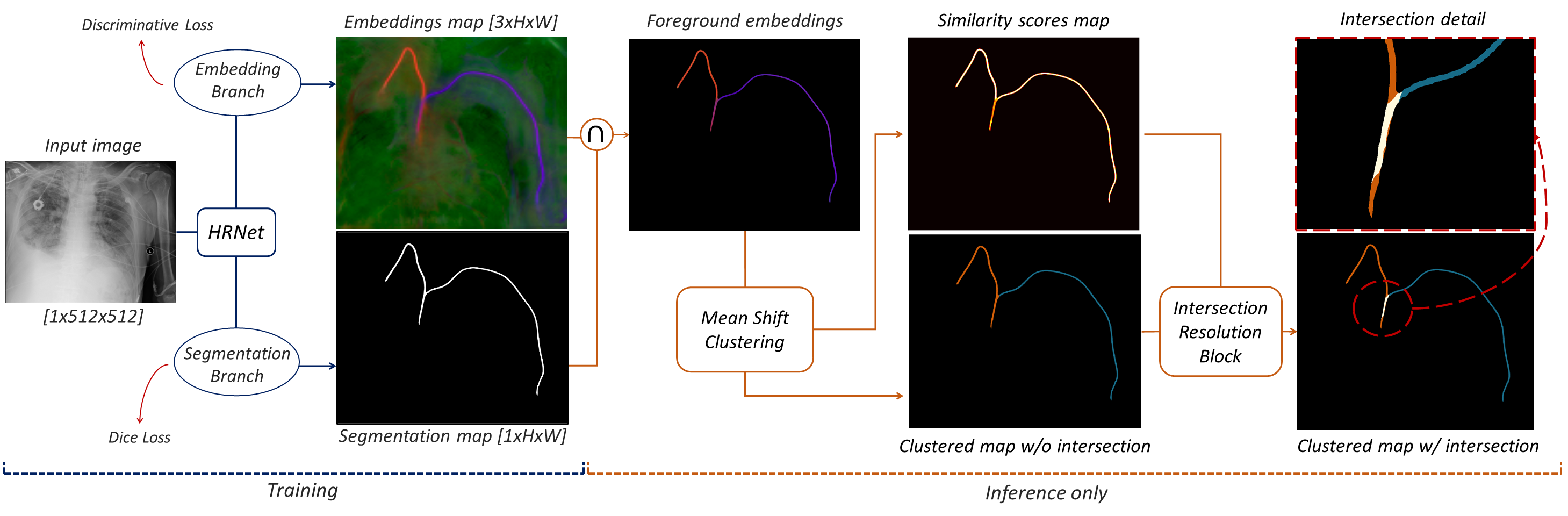}
    \centering
    \caption{Outline of our method. HRNet outputs a segmentation and an embedding map. We run mean shift clustering on predicted foreground embeddings and use the outcome to compute for each of them a similarity score $s$. The warmer the color in the similarity scores map, the smaller $s$ and the more the embedding is equally close (similar) to both cluster centers found by mean shift.
    Pixels corresponding to a similarity score lower than some threshold $a=0.7$ are classified as intersection pixels and merged with mean shift clustering results. HxW refers to the original resolution of the input image.}
    \label{fig:pipeline}
\end{figure*}

Our method was inspired by the LaneNet architecture \cite{neven2018towards}.
Originally designed for autonomous driving applications, LaneNet tackles a road lane detection task by treating it as an instance segmentation problem.
Similarly, in our study, we aim to segment cardiac catheters which are long and thin structures. 
We employ a branched multi-task network, consisting of two parts: a segmentation branch that detects the pixels belonging to a catheter, and an embedding branch that predicts an embedding value for each of them. To disentangle segmented pixels into different catheters instances, we perform clustering on their embeddings. After that, we employ an algorithm to detect intersection pixels and assign them to all the involved catheters.
Our 3-staged method can be outlined as: segmentation and embeddings prediction, embeddings clustering, and intersection resolution (Figure \ref{fig:pipeline}).\\ 
\paragraph{Segmentation and embeddings prediction.} 
In our implementation, we replaced the ENet backbone used in \cite{neven2018towards} by a HRNet architecture, which showed promising results for catheter semantic segmentation \cite{sirazitdinov2021landmark}. 
The last layer of HRNet (Figure \ref{fig:pipeline}) consists of two almost identical branches: a segmentation branch and an embedding branch, both implemented as 1x1 convolution. The segmentation branch is trained with the Dice loss to output a segmentation map. The embedding branch is trained with the discriminative loss function \cite{neven2018towards} to map each pixel in the image to a point in a 3d feature space, known as a pixel embedding. In this embedding, pixels that belong to the same catheter end up close together in the feature space, while pixels belonging to different instances lie far apart. In particular, the discriminative loss consists of a weighted sum of two terms working together (Equation \ref{eq:loss}). A variance term $L_{var}$ pulls embeddings of the same cluster towards the mean embedding (cluster center) when they are further than $\delta_v$, and a distance term $L_{dist}$ pushes different clusters further apart, increasing the space between their centers when they are closer than $\delta_d$.
\begin{equation}
\small
    \begin{cases}
L_{var} = \frac{1}{C} \sum_{c=1}^{C} \frac{1}{N_c} \sum_{i=1}^{N_c} [|| \mu_c - x_i|| - \delta_v]_{+}^{2} \\\\
L_{dist} = \frac{1}{C(C-1)} \sum_{c_{A}=1}^{C} \sum_{c_{B}=1, c_a \neq c_B}^{C} [\delta_d - || \mu_{c_A} - \mu_{c_B}||]_{+}^{2}
    \end{cases}
\label{eq:loss}
\end{equation}
In Equation \ref{eq:loss}, $C$ is the number of clusters, $N_c$ the number of elements in cluster $c$, $x_i$ a pixel embedding, $\mu_c$ the mean embedding of cluster $c$ and $[x]_+ = max(0, x)$.\\
\paragraph{Embeddings clustering.} During the inference, the network generates two outputs: a semantic segmentation map, denoting which pixels are part of a catheter and which are not, and a 3d embedding map, representing the embeddings for all pixels. Since HRNet downsamples the input size by 1/4, we upsample both maps to the original resolution using bilinear interpolation. We then extract from the embedding map only the pixels predicted as foreground by the segmentation branch. Our final 5d embeddings incorporate two additional dimensions by encoding the spatial coordinates of pixels to account for spatial information. 
Further, we use these 5d embeddings in a mean shift clustering algorithm to assign each pixel to an individual catheter.\\
\paragraph{Intersection resolution.} 
While in \cite{neven2018towards} road lanes do not intersect by definition, in our case, different catheters may be superimposed in the projection image. Using the original approach, such intersection pixels are assigned to only one catheter, resulting in an incorrect segmentation.
In order to overcome this limitation, we developed an intersection resolution algorithm. While the mean shift algorithm is a hard clustering algorithm, which simply partitions the set of foreground pixels into $n$ groups, we observed that, in the embedding space, the intersection pixels often lie between the corresponding clusters centers (catheters).
Therefore, for each pixel, we assess whether it can be identified as an intersection pixel based on a similarity score $s$, that is calculated based on the distances from the pixel embedding to all cluster centers determined using the mean shift. More specifically, assuming that mean shift predicted $n$ clusters, we compute the distances from the pixel embedding to all the clusters centers and sort them in ascending order, $d = (d_1, ..., d_n)$. Then, for each pair of distances $(d_1, d_i)$, where $d_1$ is the distance to the closest cluster center, $i \in [2, n]$, we compute
\begin{equation}
s_i = \frac{e^{-\beta * d_1}}{e^{-\beta * d_1}+e^{-\beta * d_i}}
\label{eq:sim}    
\end{equation}
If $s_i$ is lower than some threshold $a$, the pixel is considered to belong to the intersection of the $1^{st}$ and $i^{th}$ clusters (catheters), so that it is assigned to both.

\section{Experiments and results}
\label{sec:results}

\begin{figure*}[t!]
  \centering
     \centerline{\includegraphics[width=18cm]{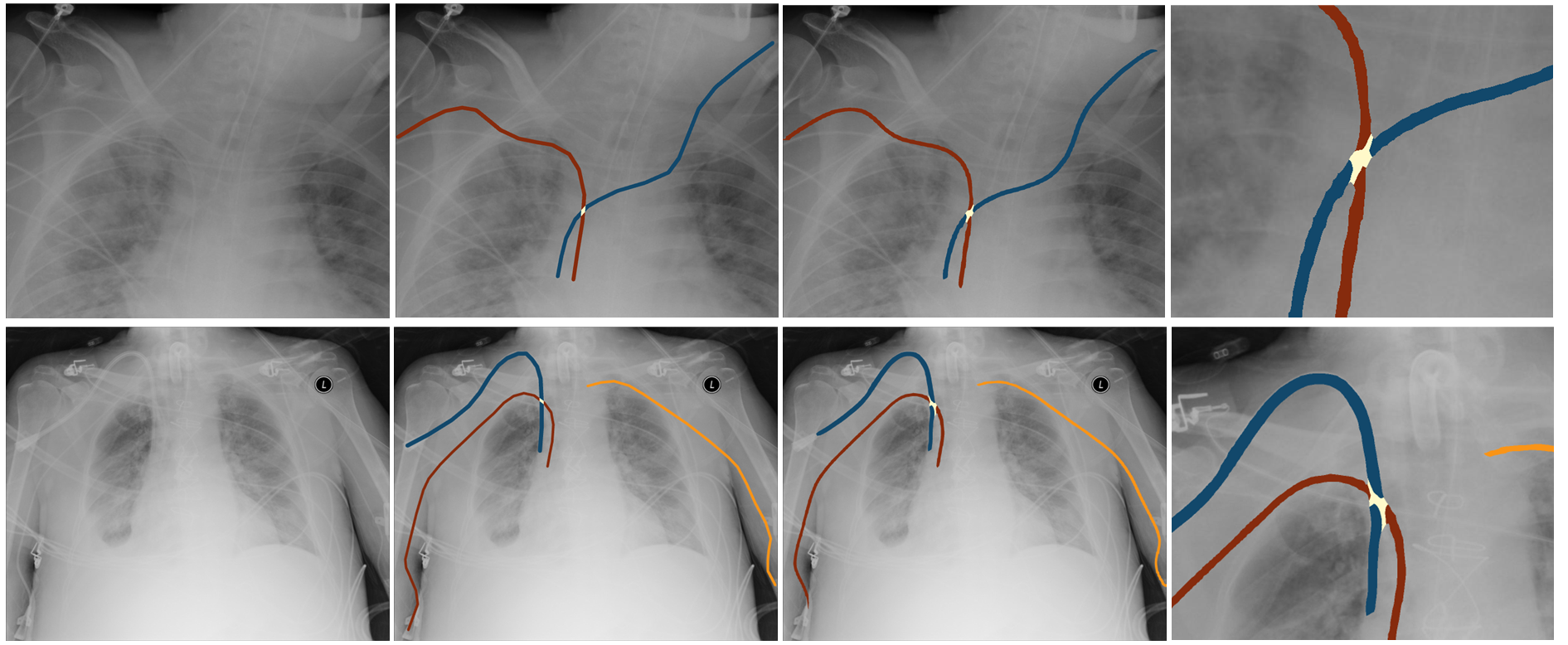}}
\caption{From left to right: CXR, ground truth, our prediction, detailed prediction and intersection area. \textit{Red}, \textit{orange}, \textit{blue} colors depicts individual instances, \textit{creamy} color shows regions assigned to multiple instances.}
\label{fig:visual_results}
\end{figure*}
In this study, we used 8877 CXRs from the RANZCR CLiP dataset \cite{tang2021clip}. The images included a total of 11786 annotated cardiac catheters, either Central Venous Catheters (CVC) or Swan-Ganz Catheters (SWG). Every CXR has at least 1 device, with 30\% of the images containing between 2 and 4 devices.
We do not differentiate between CVCs and SWG in our experiments. 
For each catheter, the dataset provides an annotation consisting of a list of pixel coordinates. We generated the segmentation annotations through a piecewise linear interpolation of the control points followed by an automated width estimation.
The linear interpolation was selected after an accurate analysis, where both linear and cubic spline interpolations were employed and visually evaluated.
We trained a LaneNet architecture with an HRNet(V2-W30) backbone pre-trained on ImageNet, with images resized to 512x512px. We performed a patient-stratified 5-fold cross-validation, using 4 folds for training (80\%) and 1 fold for test (20\%). The training part was further split into actual training data (80\%) and validation data (20\%), maintaining the patient stratification. For data augmentation the images were subjected to random rotation ($\pm 10$ degrees), horizontal flipping, as well as changes in brightness and contrast ($\pm 0.3$). Each of these transformations was applied with a probability of 20\%. The training was performed over 60 epochs, selecting the best epoch based on the lowest validation error. We used a batch size of 8 and an initial learning rate equal to 1e-4, employing AdamW optimizer with a weight decay of 0.001. The loss function was the sum of the Dice and the discriminative loss, weighted $0.3$ and $1$ respectively. During the training process, pixels in any intersection area were randomly assigned to one of the catheters involved, thereby omitting explicit modeling of the intersections. In Equation \ref{eq:sim}, we set $\beta=2$. \\
Semantic segmentation results are evaluated using Intersection over Union (IoU) and Dice coefficient, while instance segmentation is assessed using average precision (AP) and average recall (AR), averaged over IoU thresholds in a range of [0.2,0.6], with increments of 0.05. The choice of the IoU thresholds values was made according to a reader study \footnote{AP340 KI-RAD, KI-SIGS Summit 2022, Lübeck, Germany} that evaluated the quality of manual CVC annotation. The selected IoU range corresponds to an inter-reader agreement between radiologists of different levels of expertise. Results are shown in Table \ref{results}.
Figure \ref{fig:visual_results} shows successfully processed CXRs with the resolved intersected catheters. The prediction accurately matches the annotation and the intersection pixels are correctly assigned to both catheters, allowing both to be predicted as connected regions. \\
To the best of our knowledge, no previous research has been conducted using a bottom-up approach for instance segmentation of catheters in chest X-rays. To overcome the lack of a baseline from the literature, we compared our method to a semantic segmentation HRNet model, followed by morphological processing via connected component analysis (CC) on predicted binary segmentation maps to identify isolated regions. These regions become predicted instances of catheters.
Table \ref{results} shows the comparison of our model with the CC baseline. 
\begin{table}[t]
\small
\begingroup
\setlength{\tabcolsep}{4.5pt}
\renewcommand{\arraystretch}{1}
\centering
\begin{tabular}{c|c|c|c|c}
\hline
{\textbf{Model}} &{\textbf{IoU}}  &{\textbf{Dice}}  &{\textbf{AP}}  & {\textbf{AR}}\\ 
\hline 
{CC}  & {.608 $\pm$ .004} & {.746 $\pm$ .004}  & {.149 $\pm$ .013}    & {.174 $\pm$ .015} \\ \hline
{Ours}  & {.599 $\pm$ .010} & {.739 $\pm$ .009}  &
\textbf{{.726} $\pm$ {.013}}    & \textbf{{.807 $\pm$ .013}} \\ \hline
\end{tabular}
\endgroup
\caption{Comparison of semantic and instance segmentation results between our method and the HRNet semantic segmentation followed by connected components (CC) analysis. AP and AR are computed as the average over the range [0.2:0.6] with a 0.05 step of IoU thresholds.}
\label{results}
\end{table}
As it can be noticed, our method outperformed in terms of instance segmentation results, as it effectively addressed cases with catheters intersections where the connected component analysis failed, while maintaining equal performance for semantic segmentation. Additionally, we aimed to compare our model with the state-of-the-art top-down instance segmentation models. Namely, Mask R-CNN was trained and tested on the same dataset. Our attempts failed despite the careful hyperparameters search. We believe that Mask R-CNN's detect then segment approach (usually on low resolution of 28x28 or 56x56 px.) is more suited for compact objects, than for long and thin structures like catheters.
\section{Discussion and conclusion}
\label{sec:discussion}
\begin{figure}[h!]
 \centering
\centerline{\includegraphics[width=9cm]{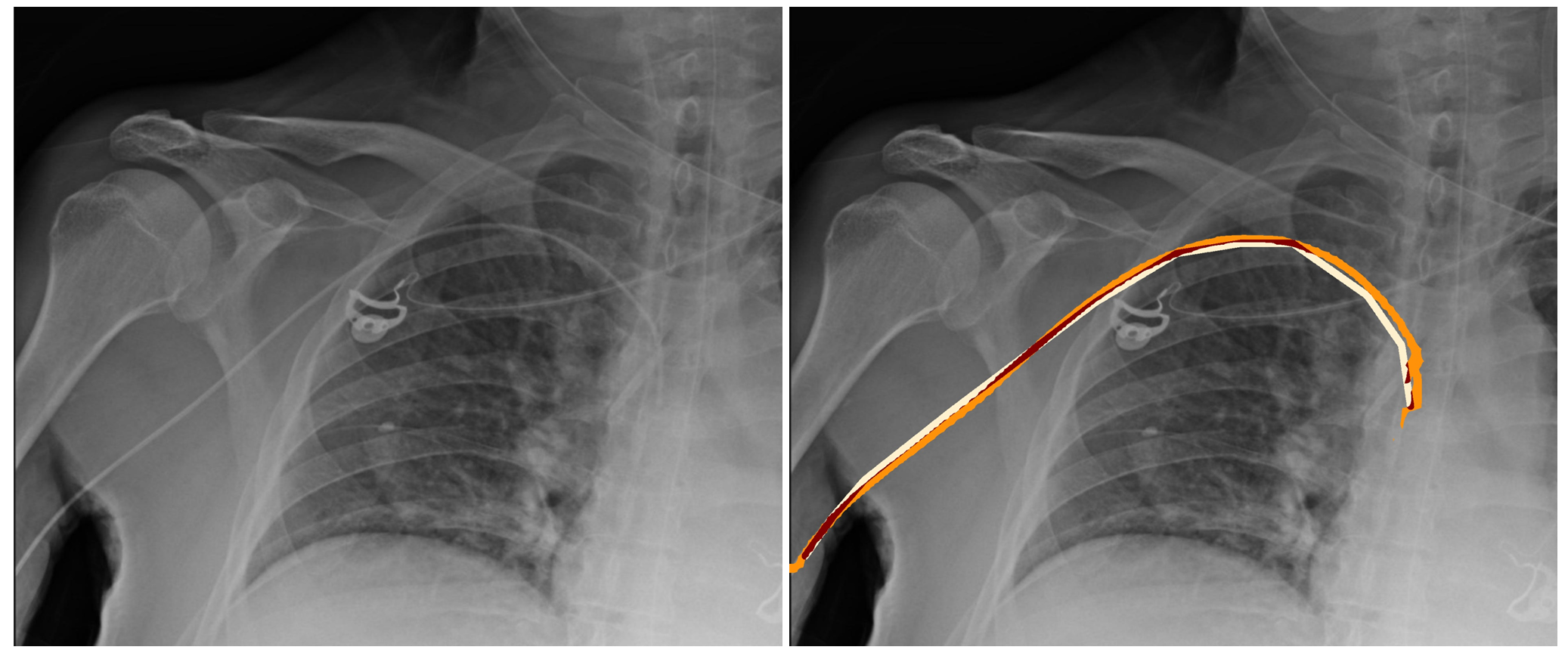}}
\caption{Overlap regions (\textit{red}) between the coarse ground truth (\textit{yellow}) and our prediction (\textit{orange}). The IoU is 0.21, despite a visually accurate result.}
\label{fig:low_iou}
\end{figure}

In this paper we presented a novel approach for the bottom-up segmentation of individual catheters in chest X-ray images. Our method is based on a branched HRNet architecture which provides next to a segmentation also an assignment of pixels to individual device instances. To the best of our knowledge, this is the first successful application of deep learning for catheter instance segmentation. 
With average AP and AR values of 0.726 and 0.807 we achieved a considerable improvement compared to a region extraction using connected component analysis. As discussed in Section \ref{sec:results}, the IoU range for calculating AP and AR values have been refined to better reflect the problem domain and the performance of radiologists in clinical practice. At the same time we observed that in several cases visually accurate predictions exhibit a low IoU value (see Figure \ref{fig:low_iou}). Such mismatches are often caused by the coarse ground truth annotation and the sensitivity of IoU to such kinds of mismatches. 
Another problem in catheter segmentation is the spatial consistency of the prediction, e.g. in the presence of multiple intersecting devices. While the proposed method copes with it employing the intersection resolution block, it is still a relevant problem that shall be addressed in future application. For instance, explicit catheter modeling as in \cite{sirazitdinov2022bi} can be adapted to instance segmentation tasks.

\bibliographystyle{IEEEbib}
\bibliography{strings,refs}

\begin{thebibliography}{10}

\bibitem{gambato2023chest}
M.~Gambato, N.~Scotti, G.~Borsari, J.~Zambon~B., J.-D. Gabrieli, A.~De~Cassai,
  G.~Cester, P.~Navalesi, E.~Quaia, and F.~Causin,
\newblock ``Chest {X-ray} interpretation: Detecting devices and device-related
  complications,''
\newblock {\em Diagnostics}, vol. 13, no. 4, pp. 599, 2023.

\bibitem{subramanian2019automated}
V.~Subramanian, H.~Wang, J.~T Wu, K.~CL Wong, A.~Sharma, and T.~Syeda-Mahmood,
\newblock ``Automated detection and type classification of central venous
  catheters in chest {X-rays},''
\newblock in {\em Medical Image Computing and Computer Assisted
  Intervention--MICCAI 2019: 22nd International Conference, Shenzhen, China,
  October 13--17, 2019, Proceedings, Part VI 22}. Springer, 2019, pp. 522--530.

\bibitem{hansen2021radiographic}
L.~Hansen, M.~Sieren, M.~Hobe, A.~Saalbach, H.~Schulz, J.~Barkhausen, and M.~P
  Heinrich,
\newblock ``Radiographic assessment of {CVC} malpositioning: How can {AI} best
  support clinicians?,''
\newblock in {\em Medical Imaging with Deep Learning}, 2021.

\bibitem{jung2022classification}
S.~Jung, J.~Oh, J.n Ryu, J.~Kim, J.~Lee, Y.~Cho, M.~S. Yoon, and J.~Y. Jeong,
\newblock ``Classification of central venous catheter tip position on chest
  {X-ray} using artificial intelligence,''
\newblock {\em Journal of Personalized Medicine}, vol. 12, no. 10, pp. 1637,
  2022.

\bibitem{he2017mask}
K.~He, G.~Gkioxari, P.~Doll{\'a}r, and R.~Girshick,
\newblock ``{Mask R-CNN},''
\newblock in {\em Proceedings of the IEEE international conference on computer
  vision}, 2017, pp. 2961--2969.

\bibitem{neven2018towards}
D.~Neven, B.~De~Brabandere, S.~Georgoulis, M.~Proesmans, and L.~Van~Gool,
\newblock ``Towards end-to-end lane detection: an instance segmentation
  approach,''
\newblock in {\em 2018 IEEE intelligent vehicles symposium (IV)}. IEEE, 2018,
  pp. 286--291.

\bibitem{sun2019high}
K.~Sun, Y.~Zhao, B.~Jiang, T.~Cheng, B.~Xiao, Y.~Liu, D.and~Mu, X.~Wang,
  W.~Liu, and J.~Wang,
\newblock ``High-resolution representations for labeling pixels and regions,''
\newblock {\em arXiv preprint arXiv:1904.04514}, 2019.

\bibitem{sirazitdinov2021landmark}
I.~Sirazitdinov, M.~Lenga, I.~M Baltruschat, D.~V Dylov, and A.~Saalbach,
\newblock ``Landmark constellation models for central venous catheter
  malposition detection,''
\newblock in {\em 2021 IEEE 18th International Symposium on Biomedical Imaging
  (ISBI)}. IEEE, 2021, pp. 1132--1136.

\bibitem{tang2021clip}
J.~SN Tang, J.~CY Seah, A.~Zia, J.~Gajera, R.~N Schlegel, A.~JN Wong, D.~Gai,
  S.~Su, T.~Bose, M.~L Kok, et~al.,
\newblock ``{CLiP}, catheter and line position dataset,''
\newblock {\em Scientific Data}, vol. 8, no. 1, pp. 285, 2021.

\bibitem{sirazitdinov2022bi}
I.~Sirazitdinov, A.~Saalbach, H.~Schulz, and D.~V Dylov,
\newblock ``Bi-directional encoding for explicit centerline segmentation by
  fully-convolutional networks,''
\newblock in {\em International Conference on Medical Image Computing and
  Computer-Assisted Intervention}. Springer, 2022, pp. 693--703.

\end{thebibliography}

\end{document}